\documentclass{llncs}

\usepackage{times}
\usepackage{subfigure}
\usepackage{color}
\usepackage{amsmath}
\usepackage{amsfonts}
\usepackage{graphicx}

\begin{document}

\pagestyle{headings}

\title{The Conceptual Integration Modeling Framework:
Abstracting from the Multidimensional Model}

\author{{\bf Flavio Rizzolo}\inst{1}\thanks{Also affiliated with Carleton University.} \and
        {\bf Iluju Kiringa}\inst{2} \and
        {\bf Rachel Pottinger}\inst{3} \and
        {\bf Kwok Wong}\inst{4}}

\institute{University of Ottawa. frizzolo@site.uottawa.ca \and
           University of Ottawa. kiringa@site.uottawa.ca \and
           University of British Columbia. rap@cs.ubc.ca \and
           University of Ottawa. kwong080@uottawa.ca}

\maketitle

\def\hiera{\preceq}

\newcommand{\eat}[1]{}
\newcommand{\tkf}{\small\sf}
\newcommand{\tkfb}{\small\sf\bfseries}

\newcommand{\blackBox}{\nobreak \ifvmode \relax \else
           \ifdim\lastskip<1.5em \hskip-\lastskip
           \hskip1.5em plus0em minus0.5em \fi \nobreak
           \vrule height0.60em width0.5em depth0.10em\fi}
\newcommand{\filledjoin}{\blacktriangleright\hspace*{-1mm}\blacktriangleleft}
\newcommand{\dotted}[1]{\buildrel\textstyle.\over{#1}}
\newcommand{\vnote}[2]{{{\blackBox}{\bf{[#1]: #2}}}}

\begin{abstract}
Data warehouses are overwhelmingly built through a bottom-up
process, which starts with the identification of sources,
continues with the extraction and transformation of data from
these sources, and then loads the data into a set of data marts
according to desired multidimensional relational schemas. End user
business intelligence tools are added on top of the materialized
multidimensional schemas to drive decision making in an
organization. Unfortunately, this bottom-up approach is costly
both in terms of the skilled users needed and the sheer size of
the warehouses. This paper proposes a top-down framework in which
data warehousing is driven by a conceptual model. The framework
offers both design time and run time environments. At design time,
a business user first uses the conceptual modeling language as a
multidimensional object model to specify what business information
is needed; then she maps the conceptual model to a pre-existing
logical multidimensional representation. At run time, a system
will transform the user conceptual model together with the
mappings into views over the logical multidimensional
representation. We focus on how the user can conceptually abstract
from an existing data warehouse, and on how this conceptual model
can be mapped to the logical multidimensional representation. We
also give an indication of what query language is used over the
conceptual model. Finally, we argue that our framework is a step
along the way to allowing automatic generation of the data
warehouse.

\end{abstract}

\section{Introduction} \label{Section:Intro}

Organizations can now access vast amounts of data drawn from a variety
of data sources including operational databases (e.g., legacy product
and service databases, financial databases, human resource or customer
databases), business documents, spreadsheets, and totally or partially
structured web documents. Data warehouses~\cite{Kimball98,Han06}
provide a physical, integrated repository of various heterogeneous
data sources.  At design time, a data warehouse schema is constructed
based on the local schemas of the data sources to be integrated.  At
run time, the warehousing process first identifies required
sources. It then successively transforms the data extracted from the
sources and materializes them into a warehouse or a set of data marts
according to a multidimensional relational schema. Finally,
applications such as cubes and various mining and modeling
tools on top of the warehouse generate business intelligence for
running the organization.

Both the design and run-time processes are bottom-up in the sense
that the warehouse's schema and data are driven only by
information to be integrated in the data warehouse, but not by the
needs of users and their Business Intelligence (BI) applications.
In the existing bottom-up methodologies, information to be
integrated into a data warehouse is the focal point of attention,
and the users' requirements are taken into account only as a
software engineering step for coming up with the conceptual model
of the warehouse, not as a schema that can be instantiated and
queried. This means that any conceptual model used is almost
exclusively for the purpose of designing the data warehouse, never
as an implementable data model against which queries are posed.

Existing query and report tools over data warehouses have tremendously
increased productivity and enabled the management of organization
performance with an ease never seen before the advent of these
tools. Using these tools to generate reports is relatively easy, and
these reports can also be easily generated on framework. However, the
difficult and open challenges lie in issues such as expressing and
executing the mappings on one hand, as well as evolving the models on
the other; this paper focuses on the former set of challenges.

\begin{figure}[t]
\begin{center}
  \begin{tabular}{| l l|| l l|}
   \hline
  \multicolumn{2}{|l|}{ \textbf{CIM Visual Model (CVM)}}  &
\multicolumn{2}{|l|}{ \textbf{CIM Data Model (CDM)}}\\
   \hline \hline
   \textbf{CVL}  &: \textbf{C}onceptual \textbf{V}isual \textbf{L}anguage & \textbf{CDL} &: \textbf{C}onceptual \textbf{D}ata \textbf{L}anguage\\ \hline
   \textbf{MVL}  &: \textbf{M}apping \textbf{V}isual \textbf{L}anguage & \textbf{MDL} &:  \textbf{M}apping \textbf{D}ata \textbf{L}anguage\\ \hline
   \textbf{SVL}  &: \textbf{S}tore \textbf{V}isual \textbf{L}anguage & \textbf{SDL} &: \textbf{S}tore \textbf{D}ata \textbf{L}anguage\\ \hline
  \end{tabular}
\end{center}
\caption{The landscape of models of the CIM Framework}
\label{fig:landscape} \vspace*{-3mm}
\end{figure}

In contrast to the existing bottom-up techniques, we introduce an
approach whose overall goal is to offer a framework for a top-down,
requirements-driven data warehouse construction. This paper describes
the Conceptual Integration Modeling (CIM) Framework, an approach that
allows users to raise the level of data integration abstraction by
specifying their needs at a high conceptual level which is then
implemented in a multidimensional platform. The CIM Framework offers
both design time and run time environments that implement a conceptual
modeling language.  This language has the two facets shown in
Figure~\ref{fig:landscape}. The first facet --- the CIM Visual Model
(CVM) --- comprises three distinct parts: (1) an extended Entity
Relationship Model~\cite{Chen76} inspired by
MultiDim~\cite{Malinowski08} and StarER~\cite{TBB99}, which are two
preexisting conceptual models for data warehouse design; (2) a visual
representation of the (relational) multidimensional model; and (3) a
visual mapping language for translating the conceptual model into the
multidimensional model.  The second facet --- the CIM Data Model (CDM)
--- is a set of XML-based, system-specific models that correspond
one-to-one to the CVM constructs. The CIM Framework
comprises the CDM and CVM as well as a query language for the
extended Entity Relationship Model above.

Our top-down integration methodology is driven by a conceptual model:
a business user specifies what business information is needed and in
what form, and the system satisfies the request.  Achieving this
requires raising the level of abstraction significantly. Today,
integration is primarily concerned with the characteristics of
individual products.  We need to move to a world where users primarily
focus on articulating their data needs in terms they understand. The
conceptual model articulates the business objectives and requirements
for the warehouse. This model can be used to both design the warehouse
and to build (populate) the warehouse. As opposed to the current
state-of-the-art static approach, the new paradigm we are introducing
offers a dynamic methodology which generates the warehouse on-demand
by using a given business conceptual model for driving the warehouse
construction and the subsequent BI stages.  As requirements evolve, so
does the conceptual model together with the warehouse design and
population mechanisms.

\begin{figure}[t]
  \centering
     \includegraphics[width=120mm]{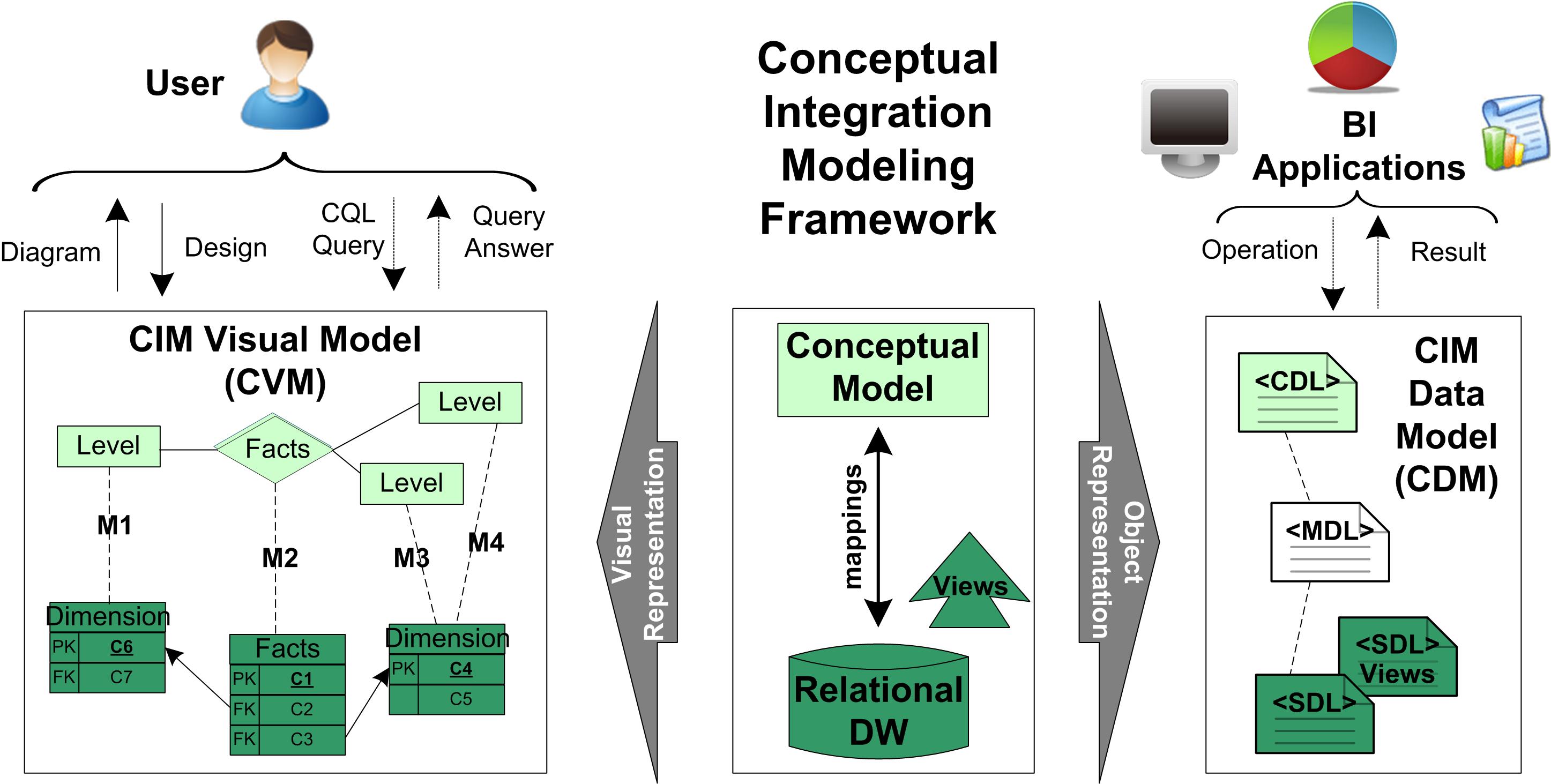}
    \caption{The various data models that are created in the CIM Framework}
    \label{fig:overall_architecture}
\vspace*{-3mm}
\end{figure}

This paper focuses on one step of our vision: how the user can
specify a conceptual model for the data warehouse, and then how
the conceptual model can be logically implemented and used in real
time. We build upon the aforementioned MultiDim and StarER to
create the CVM, and the Entity Data Model (EDM)~\cite{Blakeley06}
to define the CDM. Similarly to how EDM includes a Conceptual
Schema Definition Language (CSDL) for the conceptual level, a
Storage Schema Definition Language (SSDL) for the logical level,
and a Mapping Specification Language (MSL) to transform the
conceptual level to the logical level~\cite{Adya07}, our work
includes a Conceptual Data Language (CDL) for the multidimensional
conceptual level, a Store Data Language (SDL) for the
multidimensional (i.e., data warehouse) level, and a Mapping Data
Language (MDL) to relate the two levels. These three models form a
CIM Data Model (CDM) and are shown on the right hand side of
Figure~\ref{fig:landscape}, which is explained in detail in
Section~\ref{sec:cim}.

As shown in Figure~\ref{fig:overall_architecture}, there are
several possible paths for the user to create and manipulate data
warehouse models at design time. One path is that the user
visually creates them using the CIM Visual Models (CVM), which are
then translated into CDM specifications.  Another path is that the
user creates them using the CDM models, which then can be
visualized by CVM models for possible further manipulation. In
either case, users can query the CDL and CVL parts of the CDM and
CVM models at run time using queries formulated in a Conceptual
Query Language (CQL).  A compiler translates the MDL mappings
between CDL and SDL specifications into views over SDL, so that
the problem of processing CQL queries over CDL is reduced to a
problem of processing queries using views. This, again, parallels
the approach taken in~\cite{Adya07}, but with a much different
set of algorithms due to the different settings.

Our contributions are the following:
\begin{itemize}
\item We describe the Conceptual Integration Modeling (CIM)
Framework, which allows the top down creation of a data warehouse;

\item We describe a complete framework of languages, both visual
and data, to define conceptual and storage layers of the CIM
Framework.  While the conceptual visual language is not a major
contribution (the CVL is a variant of MultiDim), the use of such a
full suite and what it allows is. We describe the languages
themselves in enough detail to make the contributions clear;

\item We define the architecture for the CIM framework and
describe how this architecture can handle all the necessary
transformations;

\item We describe future directions, including how to extend CIM
to on-demand data warehouse creation.
\end{itemize}

The rest of the paper is organized as follows.
Section~\ref{sec:need} motivates the CIM Framework.
Section~\ref{sec:cim} describes the data-centric CVM and CDM as
depicted in Figure~\ref{fig:overall_architecture}.
Section~\ref{sec:architecture} outlines the run time architecture
that is under implementation for the CIM Framework described in
this paper. Section~\ref{sec:usage} describes some details of how
users can interact with the whole framework. We contrast our work
with related work in Section~\ref{sec:related}. Finally,
Section~\ref{sec:conclusion} concludes and describes the current
state of the implementation as well as future work to create a
variant of the data centered architecture in
Figure~\ref{fig:overall_architecture} to allow on-demand
generation of the needed parts of a data warehouse.

\section{Motivation for the CIM Model} \label{sec:need}

Almost every aspect of today's world, from business, to science,
to digital social interactions, is drowning under a deluge of
data. It is estimated that the amount of data produced in the
world grows at an astounding annual rate of
60\%~\cite{TheEconomist27022010}.  The only way to cope with this
deluge is to move toward a data-centered way of conducting human
activities --- a way that focuses on making sense of the data.
One direct consequence of the steady trend toward a data-centered
society is seen in the fact that application developers
continuously deal with managing data, which is typically
relational in format.  This is mismatched with the developers'
platforms, which are overwhelmingly object-oriented.
Additionally, the object-oriented world traditionally works at the
conceptual level, thus ensuring data independence from the
relational level, which, in many organizations, is implemented on
a variety of systems that use different representational flavors.

The divide between the conceptual and storage layer of
data-centric applications requires developing mappings between
representations to close the resulting representational gap.
Commercial (e.g., the EDM Entity
Framework~\cite{Adya07,Blakeley06}) as well open source products
(e.g., Hibernate~\cite{Hibernate}) have been proposed to bridge
the representational gap. These mapping technologies are known as
Object-Relational Mappings (ORM).

BI tools constitute another emerging class of data-centered
applications that deal with higher orders of magnitude of data
than ORM technologies. For example, Walmart operates 8,400 stores
worldwide which handle more than 200 million transactions
weekly~\cite{TheEconomist27022010}.  On one hand, the immense flow
of data that must be managed to gain insight into the retailer's
business requires using enormous multidimensional data warehouses.
On the other hand, mid-level and top executives want to make sense
of the vast amount of data at their disposal in business terms
they understand --- without having to become ultra-sophisticated
data analysts. We must bridge the gap between the world of those
executives who make decisions based on high-level,
business-oriented data representations and the world of
multidimensional models. The CIM Framework is intended to bridge
those worlds.

Many existing models (e.g., MDX and its data model ADO
MD~\cite{SpoffordHWHC06}, Universes~\cite{Howson06},
Mondrian~\cite{Mondrian}, and Framework Manager~\cite{Volitich08})
could have been used as a conceptual modeling language for
multidimensional modeling instead of our conceptual models (i.e.,
CDL and CVL). We describe them more fully in
Section~\ref{sec:conclusion}. In contrast, this section,
highlights why we do not use them as our candidate conceptual
layer.

There are two reasons not to use the four aforementioned models
for our purposes.  First, we are interested in a user-oriented,
lightweight modeling language.  None of the other candidates
displays an ease of use: for example, one can hardly imagine an
average executive who would be an expert in MDX, which is a highly
technical language whose knowledge requires a deep expertise in
multidimensional modeling. In addition, the other candidates
present a level of complexity that puts them at a level of
abstraction that is midway between the traditional
multidimensional schemas and our CIM. Second, we want a modeling
language whose constructs can be easily ascribed a clear and
declarative semantics with respect to its persistence mechanism.
We are working on such a semantics by expressing the main
components of the CDM using Datalog rules. None of the other
candidates displays a clearly defined and --- to the best of our
knowledge --- known semantics that serves as its formal
foundation. Using Datalog rules to express the foundations of the
mapping models does not conflict with the goal of a user-oriented
language since such a formalization is transparent to the users.

\section{Conceptual Integration Modeling Framework} \label{sec:cim}
\subsection{Visual Model} \label{sec:cvm}

\begin{figure}[t]
    \centering
        \includegraphics[width=120mm]{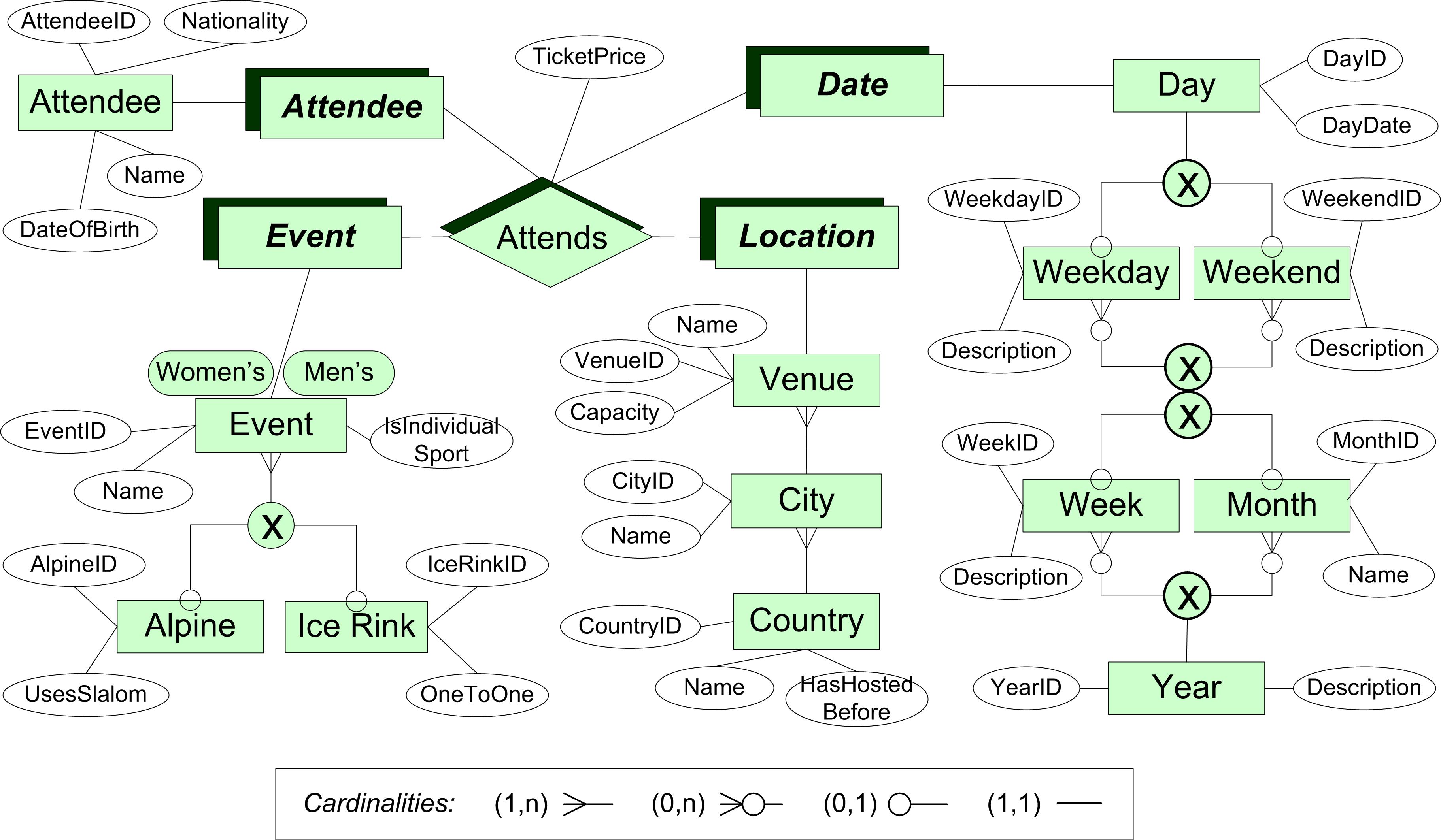}
    \caption{Olympic CVL Model}
    \label{fig:olympics}
\end{figure}

Our CVL model extends MultiDim~\cite{Malinowski08}, which in turn
extends Peter Chen's ER model to support multidimensionality. This
section both briefly reviews conceptual data modeling of
multidimensionality and introduces CVL. Readers interested in more
background than can be presented here should refer to another
source, e.g.,~\cite{Malinowski08}.

Typically, data warehouses are described and defined by a
multidimensional model in which data are points called {\it
facts\/} in an n-dimensional space called a {\it data cube\/}. The
{\it dimensions\/} of a data cube are the perspectives used to
analyze data.  A dimension consists of a set of hierarchies where
a hierarchy contains a set of {\it levels of granularity\/}. Facts
are represented by {\it fact relationships\/}, which relate
entities in different dimensions.

Throughout this section we use a CVL representation of a data
warehouse about Olympic events and attendees (See
Figure~\ref{fig:olympics}). In CVL, fact relationships are
represented by shadowed diamonds (e.g., {\tkf Attends}); shadowed
rectangles depict dimensions of the fact relationships (e.g.,
{\tkf Location}, {\tkf Event}), and levels are represented as
unshadowed rectangles (e.g., {\tkf Venue}, {\tkf City}). Unlike
MultiDim, dimensions are first-class citizens in CVL diagrams, and
thus are explicitly represented. Levels and fact relationships may
also have properties/measures, represented by ovals (e.g., {\tkf
TicketPrice}, {\tkf VenueID}). The directly attached level to a
dimension is the bottom level of that specific dimension, which
determines the granularity of the facts in the respective fact
relationship. For instance, the {\tkf TicketPrice} values in {\tkf
Attends} are given by {\tkf Venue}, {\tkf Day}, {\tkf Event} and
{\tkf Attendee}.

A dimension may represent more than one hierarchy, so we introduce
a flattened oval to denote the specific hierarchy (e.g., {\tkf
Women's} and {\tkf Men's}). If a dimension has only one hierarchy,
or all hierarchies in a domain contain the same levels, we omit
the hierarchy oval.  For example, each of the {\tkf Location},
{\tkf Date} and {\tkf Attendee} dimensions contain only one
hierarchy, so a separate hierarchy oval is omitted.

Instances of levels are related hierarchically, unlike the flat,
two way relationships in ER diagrams.  For example, {\tkf City} is
a child of {\tkf Country}. A hierarchy may have different options
for a level of granularity. For example, a {\tkf Day} may roll up
to either a {\tkf Weekday} or a {\tkf Weekend}, but not both.
Similarly, a {\tkf Year} may drill down to exclusively a {\tkf
Week} or a {\tkf Month}. This so called splitting/joining
parent-child relationships are expressed with parent-child
relations labeled by an encircled '{\tkfb X}'. In contrast to
MultiDim, which has splitting and joining levels instead of
parent-child relationships, the '{\tkfb X}' CVL relationships can
be combined. This is exemplified by the double '{\tkfb X}'
connecting {\tkf Weekday} and {\tkf Weekend} to {\tkf Week} or a
{\tkf Month}, representing four mutually exclusive parent-child
relationships: {\tkf Weekday}--{\tkf Week}, {\tkf Weekend}--{\tkf
Week}, {\tkf Weekday}--{\tkf Month}, and {\tkf Weekend}--{\tkf
Month}. This is different from a disjoint inheritance in Chen's
ER, where entities can have different types and one entity cannot
inherit from two entities.

\begin{figure}[t]
    \centering
        \includegraphics[width=100mm]{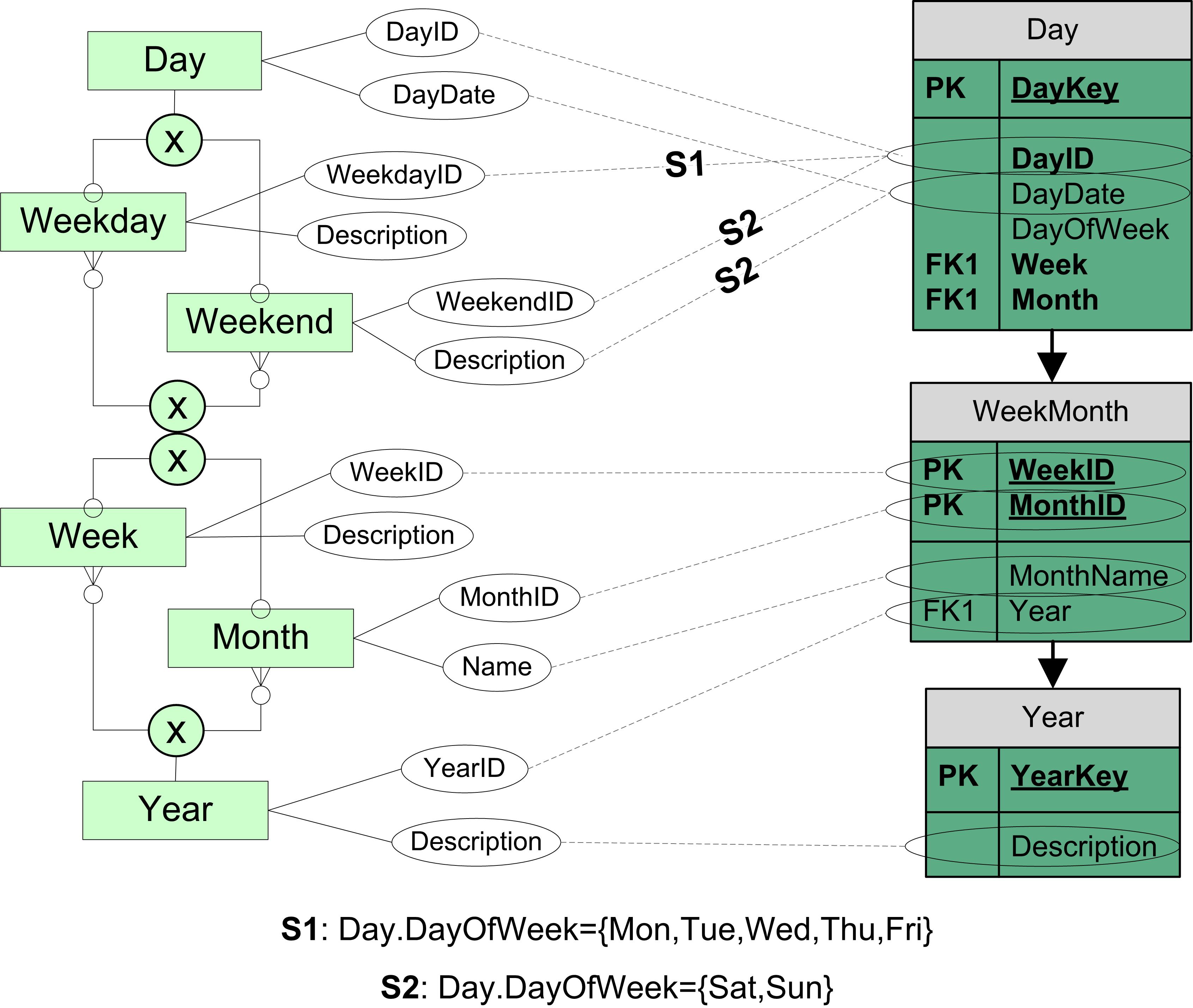}
    \caption{CVM model of a portion of the Olympic example}
    \label{fig:olympics_cvm}
\end{figure}

Parent-child relationships have cardinality constraints indicating
the minimum and maximum number of members in one level that can be
related to a member in another level. For example, {\tkf City} is
related to {\tkf Country} with a {\tkf (1,n)} to {\tkf (1,1)}
cardinality: every city belongs to a country and each country has
many cities.

The SVL is a UML-like representation of the relational data
warehouse schema, containing relational table definitions, keys
and referential integrity constraints. In practice, the SVL
specification is automatically generated from the underlying data
warehouse schema, and is put at the user's disposal.

In addition to the CVL and SVL specifications, the user also
provides visual mappings between constructs in both models; the
mappings constitute an MVL specification.
Figure~\ref{fig:olympics_cvm} shows a CVM model of a portion of
the Olympic example: the left hand side of the figure shows a
subset of the CVL and the right hand side depicts part of the SVL
of a pre-existing Olympic data warehouse. Between the CVL and the
SVL specifications, mappings in the form of dotted lines form the
MVL specification. The CVL specification corresponds to what is
increasingly called the {\it semantic layer} in industry. Such a
layer liberates users from the low-level multidimensional
intricacies and allows them to focus on a higher level of
abstraction. For instance, the SVL model for the logical data
warehouse on the right hand side has normalized tables for {\tkf
Day} and {\tkf Year} but a de-normalized joint table for {\tkf
Week} and {\tkf Month}. This mix of normalized and de-normalized
warehouse schemas are common in real-world applications. In
contrast, the CVL model on the left hand side has separate
entities for {\tkf Week} and {\tkf Month} and two additional
entities that are not even represented in the logical data
warehouse schema: {\tkf Weekday} and {\tkf Weekend}. This divide
between the conceptual and the logical levels is bridged by the
MVL.

The MVL can also contain conditions to select data from an SVL
table. For instance, conditions {\tkfb S1} and {\tkfb S2}
determine what data from the {\tkf Day} table is mapped to {\tkf
Weekday} and {\tkf Weekend}, respectively. MVL also allows for a
conceptual entity to span more than one logical table, e.g., {\tkf
Year} has one attribute mapped to the {\tkf WeekMonth} table and
another to the {\tkf Year} table. Note that not all attributes
need to be mapped; attributes like {\tkf Description} in {\tkf
Weekday} and {\tkf Week} can be mapped later to the same or other
data sources, or can be completed directly by the user.

\subsection{Data Model}
\label{sec:cdl}

\begin{figure}[t]
    \centering
        \includegraphics[width=120mm]{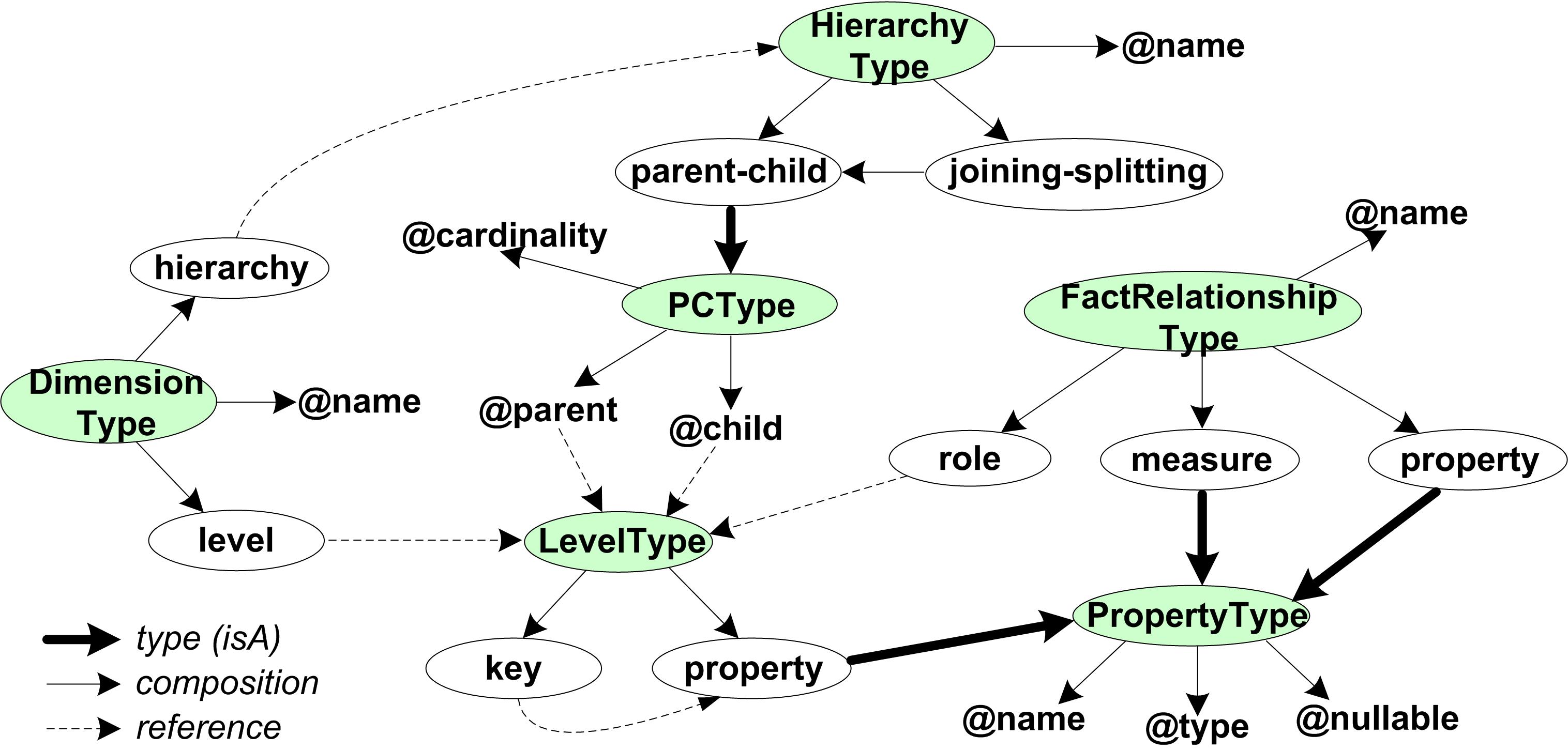}
    \caption{CDL types}
    \label{fig:csdl}
\end{figure}

The CIM Visual Model provides a simple way for the user to model
the concepts of interest and their mappings to the logical data
warehouse at hand. To enable interoperability with BI
applications, each visual component --- CVL, MVL, and SVL --- is
translated into an equivalent XML-based object model --- CDL, MDL,
and SDL, respectively. In this XML representation, each construct
is described by an XML Schema type. This separation of the CDL,
MDL, and SDL layers parallels similar layers in the EDM
Framework~\cite{Blakeley06}. However, whereas EDM provides
structure for the relational database model, our CDM provides
structure for the relational multi-dimensional model. Thus, a user
can design concepts in CVM that are automatically translated into
CDM for run-time usage.

\begin{figure}[t]
    \centering
        \includegraphics[width=120mm]{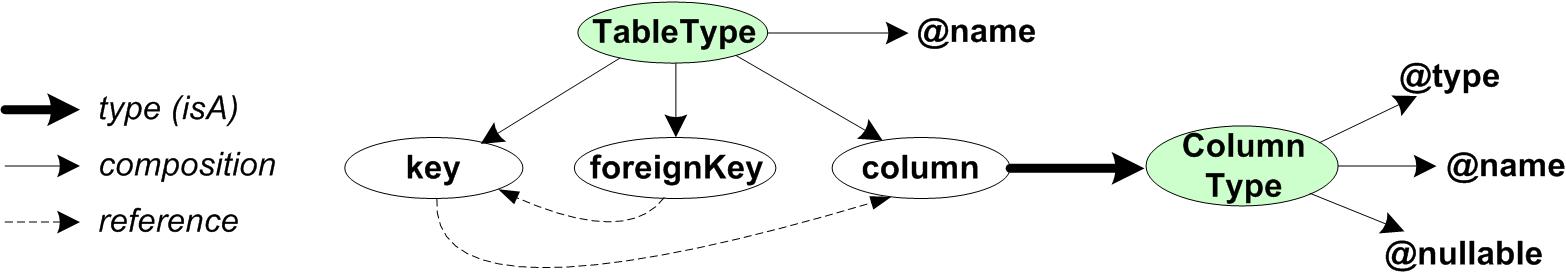}
    \caption{SDL types}
    \label{fig:ssdl}
\end{figure}

Figure~\ref{fig:csdl} shows the CDL XML Schema types definitions
and their interrelations. The green ovals are the roots of each
type, the white ovals represent the elements composing the types,
and attributes are prefixed by a ``{\tkf @}''. There is one type
for each main construct in the CVL. As usual, a composition (solid
lines) indicate nested elements in the XML instance, e.g., a
{\tkfb FactRelationshipType} has nested {\tkf role}, {\tkf
measure} and {\tkf property} elements. IsA relationships (bold
lines) provides typing information for an element, e.g., the {\tkf
property} element within a {\tkfb LevelType} is of type {\tkfb
PropertyType}. References (dashed lines) are referential
constraints expressed by \emph{key} and \emph{keyref} XML Schema
constraints. Although references are defined between {\tkf @name}
attributes, in the figure we link the respective elements instead
for clarity. A key constraint guarantees that the attribute in its
definition has a unique value within the scope in which the key is
defined. For instance, the reference between {\tkf key} and {\tkf
property} elements within a {\tkfb LevelType} forces each key
value to refer to a property within the same level.

The root of a CDL model is a complex type that contains required
{\tkf levelSet}, {\tkf dimensionSet} and {\tkf
factRelationshipSet}, and optionally may include a {\tkf
hierarchySet}. Each of these sets contains a list of all
constructs that appear in a given model (levels, dimensions, fact
relationships and hierarchies, respectively).

The SDL layer defines how the warehouse is physically stored. It
builds on the relational model and thus has two XML Schema types:
{\tkfb TableType} and {\tkfb ColumnType}. As usual, each
relational table has a primary key and may have a set of foreign
keys to other tables. The SDL model root is a complex type
containing sets of fact tables and dimensions. Each set contains a
list of all constructs in a given model. Figure~\ref{fig:ssdl}
shows the SDL XML Schema types definitions and their
interrelations.

The MDL layer expresses mappings between CDL and SDL elements. The
attribute-to-attribute correspondences provided by the users in
the MVL are grouped into \emph{mapping fragments}. A mapping
fragment contains all attribute associations between one CDL
entity and one SDL table. A mapping fragment is represented in MDL
as a {\tkf level-mapping} or {\tkf factrel-mapping} element
containing the names of the CVL entity and the SVL table being
associated. Each mapping fragment contains a set of one or more
{\tkf property-mapping} elements and optional {\tkf condition}
elements. Each {\tkf property-mapping} element represents one of
the user-defined associations between a CVL property and an SVL
table column. The {\tkf condition} elements specify the attribute
values for which the mapping fragment holds.

\begin{figure}[t]
    \centering
\subfigure[CDL Weekend
level]{\label{fig:CDLweekend}\includegraphics[width=95mm]{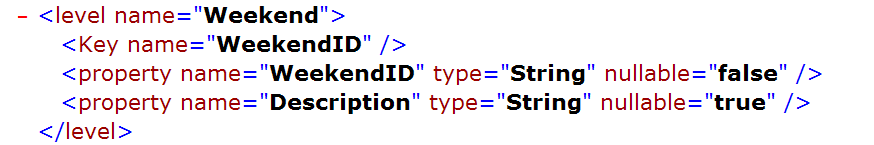}}
\subfigure[SDL Day
table]{\label{fig:SDLday}\includegraphics[width=95mm]{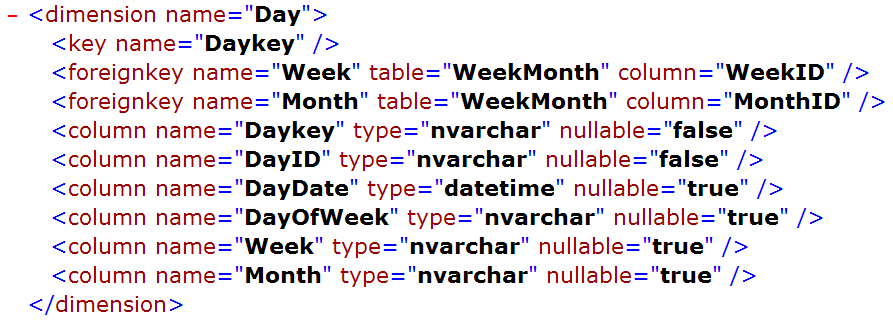}}
\subfigure[MDL Weekend-to-Day
mapping]{\label{fig:MDLweekendtoday}\includegraphics[width=95mm]{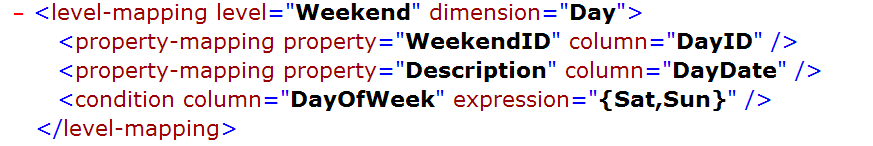}}
    \caption{CDM level instance and \textbf{S2} mapping for the Olympic example}
    \label{fig:olympics_cdm}
    \vspace*{-3mm}
\end{figure}

Figure~\ref{fig:olympics_cdm} shows a fragment of the CDM
specification for the CVM example in
Figure~\ref{fig:olympics_cvm}. Figure~\ref{fig:CDLweekend}
and~\ref{fig:SDLday} contain the definition of the {\tkf Weekend}
level in the CDL and the Day table in the SDL, respectively.
Figure~\ref{fig:MDLweekendtoday} shows the definition of mapping
{\tkfb S2}, which includes the condition that the value of the
{\tkf DayOfWeek} column has to be either {\tkf ``Sat''} or {\tkf
``Sun''}.

It is important to note that our solution requires the user to
only provide very simple property mappings between attributes in
both models, which are later transparently compiled by the system
into complex, fully-fledged, multidimensional mappings that can be
used for query evaluation.

\section{Architecture} \label{sec:architecture}

\begin{figure}
 \vspace*{-3mm}
    \centering
        \includegraphics[width=90mm]{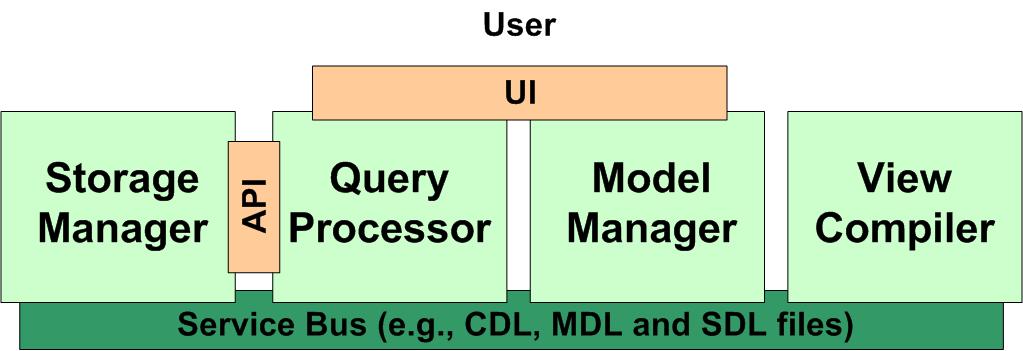}
    \caption{CIM Framework functional architecture}
    \label{fig:architecture}
\vspace*{-3mm}
 \end{figure}

We can now describe the architecture of the CIM Framework as it is
being currently implemented. The main components of this
architecture are given in Figure~\ref{fig:architecture}.

The functionality of the components is described as follows:

\begin{itemize}
\item {\bf Storage Manager}. This component creates the SVL models
automatically from the metadata exported from the underlying data
warehouse. Currently, we use Mondrian~\cite{Mondrian}, an open
source ROLAP engine, for this purpose. The component also
maintains the SDL views created by the View Compiler component
together with any other materialized and virtual view created in
the system, either relational or multidimensional.

\vspace{2mm}

\item {\bf Query Processor}. The query processor rewrites user
queries posed on the CDL model in terms of the SDL views created
by the View Compiler and sends the rewritten query to the Storage
Manager. This component treats the processing of queries as an
instance of the classical problem of processing queries using
views.

\vspace{2mm}

\item {\bf Model Manager}. This component takes a CVM model as input (i.e.,
the SVL model generated by the Store Manager, and the CVL and
MVL models provided by the user). The component produces a CDM model
(i.e., CDL, MDL, and SDL specifications) as output. Any changes in the CVM
model by the user will be incrementally propagated to the CDM.

\vspace{2mm}

\item {\bf View Compiler}. This component takes the CDL, MDL, and
SDL models generated by the Model Manager, and produces
multidimensional views over the SDL model. The component creates a
view definition for each fact relationship, level and parent-child
relationship in the CDL based on the mappings (MDL) and the
referential constraints that appear in the SDL. It may also
restructure the hierarchies into summarizable ones~\cite{HM01} if
needed. These views allow queries posed against a CDL schema to be
rewritten and posed against the compiled SDL views.
\end{itemize}

Users interact with the Model Manager and the Query Processor
components via a graphical User Interface. All modules communicate
with each other via a service bus. For the first prototype under
implementation, this bus is simply the operating system file
system where the models are stored and read in the form of XML
files. We plan to replace the file system by web services in a
service oriented architecture. Finally, the Query Processor
component communicates directly with the Storage Manager component
via a well defined application programming interface.

\section{Sample Usage Scenario}
\label{sec:usage}

Section~\ref{sec:cim} described the major components of the CIM
Framework. This section gives an overview of a usage scenario by
briefly describing how a user creates CVL and CDM models and uses
them at run time to answer queries.

\begin{figure}[h]
\vspace*{-4mm}
    \centering
    \includegraphics[width=110mm]{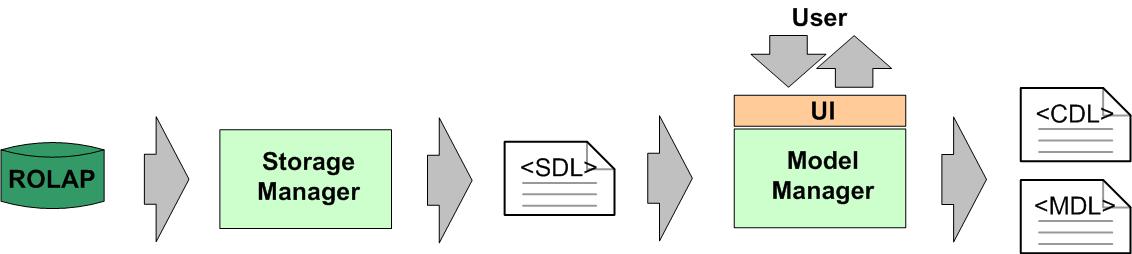}
    \caption{Model creation dataflow}
    \label{fig:modelcreation}
\vspace*{-4mm}
\end{figure}

\paragraph{Model creation.}

Figure~\ref{fig:modelcreation} outlines the first step in the
usage of the CIM Framework. The user --- typically, a business user
--- specifies a CVL model of the data warehouse (as described in
Section~\ref{sec:cvm}) using a graphical user interface offered by
the Model Manager. Then, using the Model Manager, the user imports
the SDL that represents the logical multidimensional schema of an
existing data warehouse. The Storage Manager produces such an SDL
model from that existing multidimensional schema. The imported SDL
will be visualized as an SVL model for further usage. Furthermore,
using the Model Manager graphical user interface, the user
establishes an MVL model by mapping the CVL constructs to the SVL
constructs. Finally, the Model Manager produces the CDL and MDL
models corresponding to the input CVL and MVL models,
respectively. Figure~\ref{fig:olympics_cvm} illustrates a portion
of the CVM corresponding to the Olympic example of
Figure~\ref{fig:olympics}. It is important to notice that,
similarly to the classical data independence ensured by the
distinction between the conceptual and the logical layers in
relational databases, the use of mappings in our framework ensures
an independence of the user-defined CVL and CDL layers from the
underlying multidimensional layer of the data warehouse.

\begin{figure}[h]
    \centering
    \includegraphics[width=110mm]{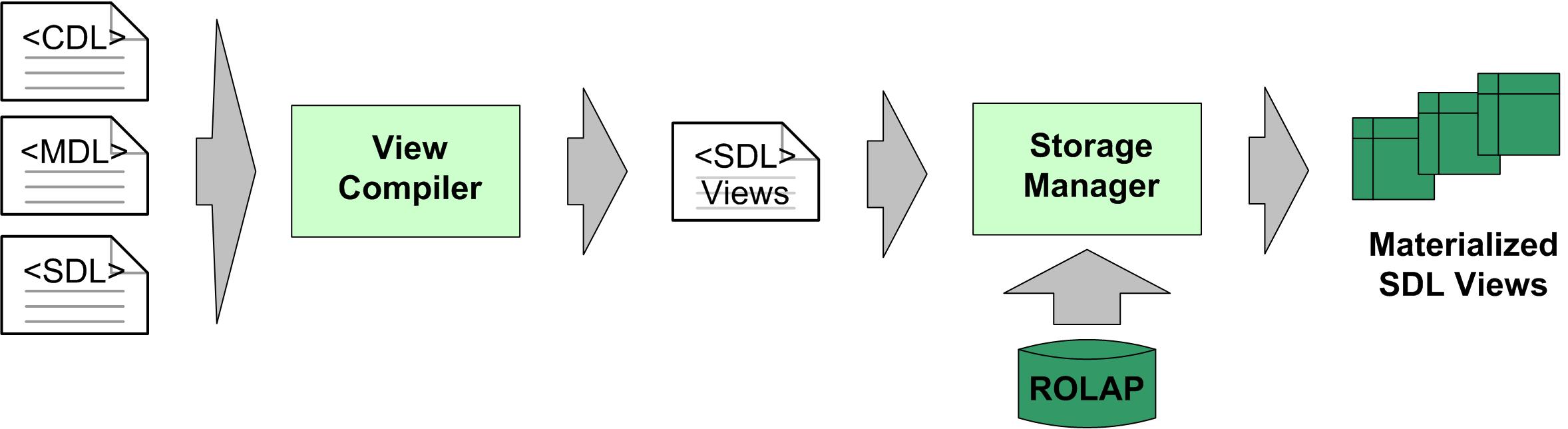}
    \caption{View compilation dataflow}
    \label{fig:viewcompilation}
\vspace*{-4mm}
\end{figure}

\paragraph{View compilation.}

Once the CDM model (i.e., the CDL, MDL, and SDL models) has been
generated by the Model Manager, a view compiler takes this CDM
model as input and produces SDL view definitions, each of which is
maintained -- either materialized or virtual -- by the Storage
Manager. CIM uses mappings of the common form $c \rightsquigarrow
\psi_{_\mathcal{S}}$, where $c$ is an element in the \emph{target}
schema and $\psi_{_\mathcal{S}}$ is a view over the \emph{source}
schema. In CIM, the conceptual model (CDL) functions as the target
schema and the store model (SDL) as the source schema. The element
$c$ in the mapping expression is either a level, a parent-child
relationship, or a fact relationship, whereas
$\psi_{_\mathcal{S}}$ is a view over the data warehouse's
multidimensional tables. CIM uses \emph{sound} mappings, i.e.,
those defined by $\forall \bar{x}(\psi_{_\mathcal{S}}(\bar{x})
\rightarrow c(\bar{x}))$ expressions, where $\bar{x}$ denotes a
set of variables. Figure~\ref{fig:viewcompilation} depicts the
compilation process. Here, we assume a ROLAP data warehouse.

\paragraph{Query formulation and evaluation.}

Consider again the Olympic example in Figure~\ref{fig:olympics}.
Suppose the user now wants to aggregate the ticket sales in the
``Attends'' fact relationship by weekend and for one specific
venue: ``Whistler Olym\-pic Park''. This query would be expressed
in the CVL query model UI with an \emph{aggregated fact
relationship} construct, ``Aggr\_Attends'', that has links to the
respective fact relationship (``Attends'') and the levels being
aggregated (``Weekend'' in Date, and ``Venue'' in Location). Since
the user wants the aggregation for a single venue, the link from
``Aggr\_Attends'' to ``Venue'' will be labeled by the selection
condition, in this case ``Venue.name=Whistler Olympic Park''. All
levels that are not being aggregated or filtered by any value
condition, such as ``Attendee'' and ``Event'', do not have to be
specified in ``Aggr\_Attends'' --- they can be obtained from the
respective fact relationship. In addition, each aggregated fact
relationship is linked to another construct specifying the
aggregation function and measure, in this case ``sum'' and
``TicketPrice''.

\begin{figure}[h]
\vspace*{-2mm}
    \centering
    \includegraphics[width=90mm]{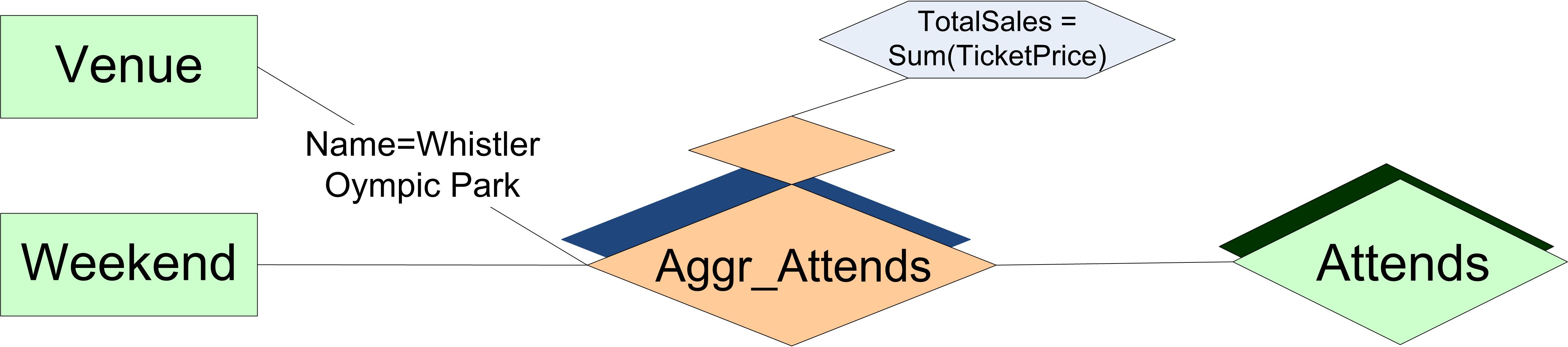}
    \caption{Aggregation query for the olympics example.}
    \label{fig:query}
\vspace*{-4mm}
\end{figure}

\begin{figure}[h]
\vspace*{-2mm}
    \centering
    \includegraphics[width=80mm]{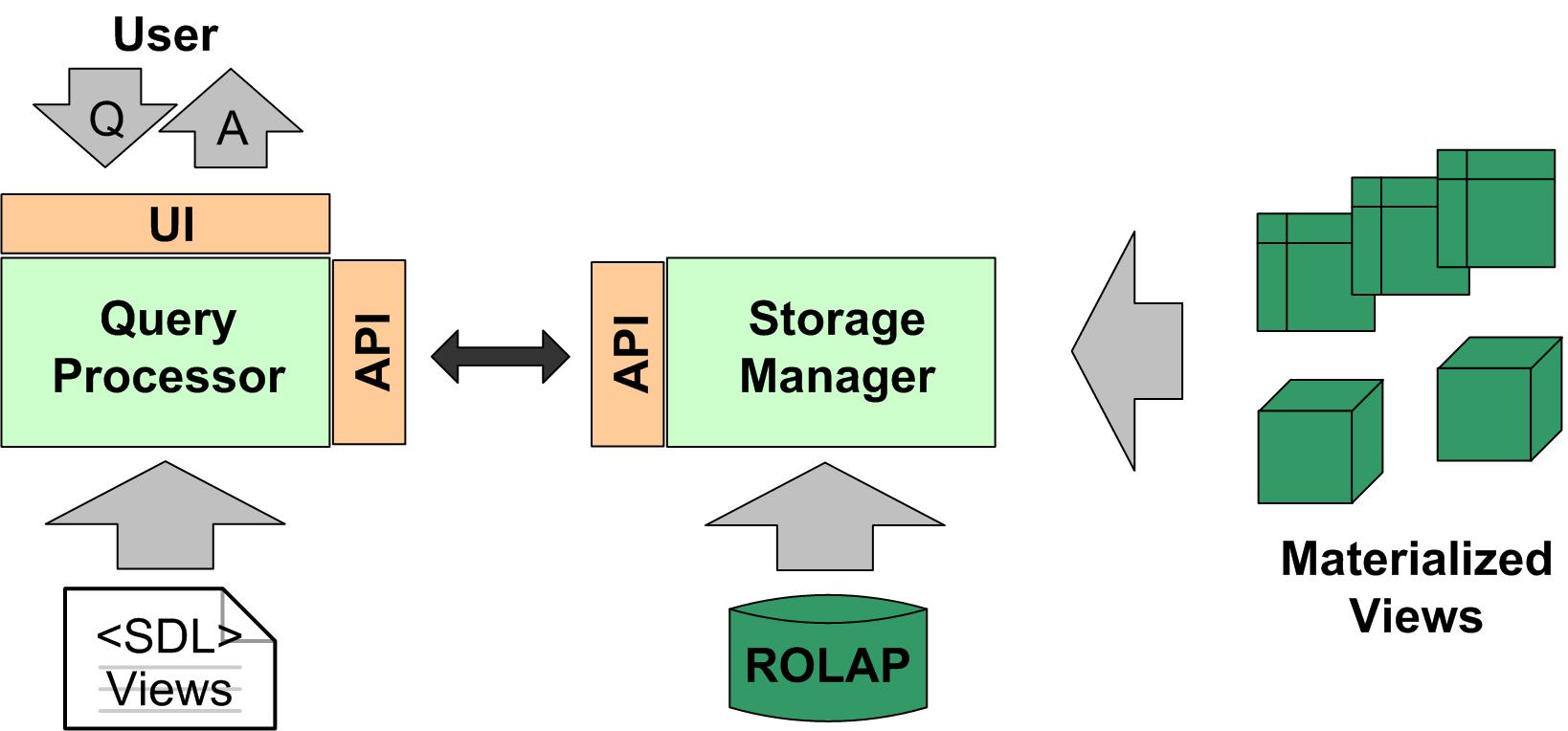}
    \caption{Query evaluation dataflow}
    \label{fig:queryevaluation}
\vspace*{-4mm}
\end{figure}

\noindent Once the query is posed to the Query Processor (QP) via
the UI, the QP uses the SDL view definitions to rewrite the query
in terms of the views. For this example, one of the view
definition would come from the S2 mapping in
Figure~\ref{fig:olympics_cvm} which selects all tuples in ``Day''
table with ``DayOfWeek=\{Sat,Sun\}'' and is used during View
Compilation to define an SDL view for weekend days. The rewritten
query is then sent to the Storage Manager (SM), which process the
query using the compiled SDL views, any precomputed cubes at hand
and the data warehouse base data. The answer is then returned to
the QP and finally to the user via the UI.
Figure~\ref{fig:queryevaluation} depicts the query evaluation
process.

\section{Related Work} \label{sec:related}

 Warehousing strategies have been classified into the
 following three categories~\cite{List02}: data-driven, goal-driven, and
 user-driven. Data-driven methodologies use the data models of
 the production database systems as the departure point of the modeling
 requirements of the data warehouse. Goal-driven methodologies use the
 corporation business processes and requirements (goals) to systematically
 derive the model of the data warehouse. Finally, user-driven methodologies
 derive the data warehouse requirements from  the user in the first place.
 As already mentioned above, all these categories of current strategies
 consider information to be integrated as the focus of the data warehousing
 process and any  goal or user-driven aspects are considered
 mostly once the integration is performed. Thus, as such, they still are
 largely bottom-up.

 We believe that, while data-driven methodologies are inherently bottom-up by
 nature, the goal and user-driven methodologies form one single category of
 stakeholder-driven methodologies, since users are a subset of
 stake holders which also comprise organizations.  Stakeholder-driven
 methodologies are inherently top-down since they free the stake holders
 as much as possible from the burden of taking care of low level tasks.
 Moreover, it has been noticed in~\cite{Kimball98} that the bottom-up
 strategy for building the warehouse is contrasting with a top-down strategy
 which the authors doubt could be possible; they even go as far as to
 assert that using some form of conceptual model to that end is not feasible.
 Our CIM Framework is a challenge to this doubt: we are building a
 tool showing that a top-down approach is possible where
 stakeholder requirements are specified in some given modeling language
 first, and then the data warehouse design and subsequent population are
 derived from those specified requirements.

Conventional database design includes the three well-known phases of
conceptual design (using the Entity Relationship model
originally proposed by Peter Chen~\cite{Chen76}),
logical design, and physical design. These phases result
in the creation of database schemas at each of the conceptual, logical, and
physical levels. Contrary to the assertion in~\cite{Kimball98}, the last ten
years have witnessed the emergence of some approaches for data warehouse
design which adapt the design phases used in the conventional design of
operational databases (including the conceptual design phase) to the
characteristics of data warehouses~\cite{Torlone03,Malinowski08}.
One such characteristic is the consideration of data provenance and
availability in data warehouse conceptual models.

Despite some progress in the last ten years with respect to
multidimensional modeling, the field of conceptual modeling of data
warehouse is still in its infancy as recognized in~\cite{Malinowski08}.
As noticed in~\cite{Malinowski08}, many of the approaches taken towards
multidimensional modeling are at the logical level. Many others
restrict themselves to a subset of the features of a data warehouse.
Only a few have tackled the issue of conceptual multidimensional
modeling by taking into account all facets of a data
warehouse~\cite{Malinowski08}.

Conceptual models for data warehouse design can be broadly
classified into graphical~\cite{Hahn00,SBHD98,ASS06,Golfarelli98}
and non-graphical~\cite{PJD01} approaches. We restrict our
attention to graphical approaches.  There are a number of
different graphical representations of conceptual multidimensional
schemas for warehouse modeling,
e.g.,~\cite{SBHD98,Golfarelli98,Cabibbo98,Hahn00}.  Hahn {\em et
al.} describe the design and implementation of a tool for
automatically generating On Line Analytical Processing (OLAP)
schemas from conceptual graphical models~\cite{Hahn00}.  They use
a multidimensional entity relationship (M E/R) model as the
conceptual graphical language; M E/R is a extension of the ER
model~\cite{Chen76} for multidimensional purposes.  Malinowski and
Zim\`{a}nyi describe a multidimensional model called MultiDim for
representing OLAP and data warehouse
requirements~\cite{Malinowski08}.  The model represents the
various data warehouse concepts such as facts, dimensions,
hierarchies, measures, etc., by means of a graphical notation
similar to the ER model. Moreover, the model puts a special
emphasis on capturing the various sorts of (often irregular)
hierarchies that appear in real-world applications. Malinowski and
Zim\`{a}nyi also provide mapping rules for implementing the
MultiDim schemas into the object-relational logical model. This
mapping is based on the classical rules for mapping ER models to
the relational model.

The graphical approaches mentioned above are mainly proposals for
modeling languages made as part of data warehouse design
methodologies. They are rarely used as a run time environment. By
contrast, our proposal provides a run-time environment that allows
a user to pose queries and do business analytics at a conceptual
level that is abstracted from the logical multidimensional data
level.

As already mentioned in Section~\ref{Section:Intro}, the EDM
Framework~\cite{Adya07,Blakeley07,Blakeley06} provides a querying and
programming platform that raises the level of abstraction from the
logical relational data level to Peter Chen's ER conceptual level.  At
design time, a developer provides three artifacts: an Entity
Relationship schema written in a conceptual schema definition
language (CSDL), a relational database schema expressed in a store schema
definition language (SSDL), and a mapping between the CSDL and SSDL schema
elements. These three components make up an EDM model. In addition, a
query language, Entity SQL (eSQL), is defined over the CSDL. A
compiler takes an EDM model as input and generates the mapping
information in the form of eSQL views that express ER constructs in
terms of relational tables. At run time, eSQL queries over the CSDL
are run against the eSQL views via view unfolding.

Hibernate is an Object-Relational technology that
allows the creation of persistent Java classes. At design time,
the developer writes a Java program that comprises classes and
mappings of those classes to SQL queries over an underlying
relational schema. The mappings involved here are not compiled at
all: they solely serve the purpose of  directly translating
objects to SQL queries.

Mondrian~\cite{Mondrian} is an open-source OLAP engine implemented
with Java technology. It executes queries written in MDX over data
coming from a relational database backend (e.g., MySQL,
PostgreSQL, Oracle), and presents the results in a
multidimensional format via a Java API. Mondrian's goal is to
provide data warehouse and OLAP functionality on top a relational
DBMS.

Like our CIM Framework, EDM and Hibernate aim to bridge the gap
between the prevalent object-oriented world of application
programmers and the world of relational data. Unlike our CIM
Framework, which deals with the multidimensional data model, EDM
and Hibernate deal with the classical relational data model.

On the industry side, two major vendors (namely SAP Business
Objects and IBM) of business analytics solutions provide
proprietary conceptual levels that they call ``semantic layers''.
SAP Business Objects' semantic layer~\cite{Howson06}, called a {\it
Universe}, is a business representation of an organization's data
asset (i.e., data warehouse as well as transactional
databases). The universe lies between the end user and the
organization's data asset; it hides the complex structure of the
underlying data assets as well as their provenance and
localization from the end user by providing the later with a
familiar business-oriented interface. IBM's semantic layer,
Framework Manager~\cite{Volitich08}, is similar to SAP's
Universes and works according to the same principles.

\section{Conclusion and Current Status of the Implementation}
\label{sec:conclusion}

 \begin{figure}[t]
    \centering
        \includegraphics[width=120mm]{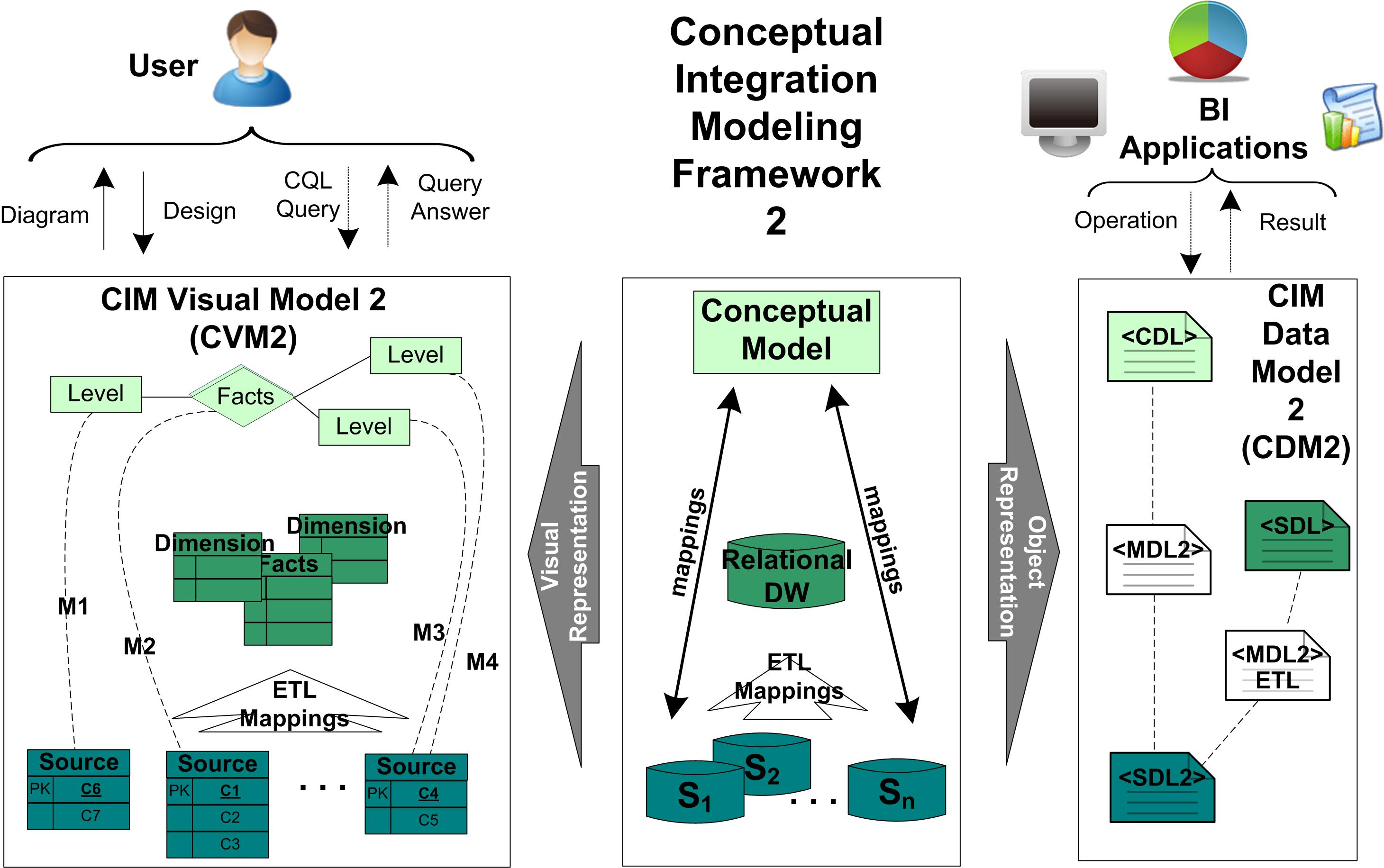}
    \caption{Data centered architecture for automatic generation of the data warehouse}
    \label{fig:overall_architecture2}
 \end{figure}

An increasing number of business analytics application are being
used by business executives and end users who are infrequently
schooled in the intricacies of data warehousing, data integration,
and complex query formulation technologies. Technologies are
emerging in Industry to tackle the issue of bridging the gap
between business decision applications, which are end-user
oriented, and the existing, complex data warehousing technology
via the use of abstract layers. The CIM Framework fills a void in
academia for studying these abstract layers in a principled way.
We are currently implementing the architecture depicted in
Figure~\ref{fig:architecture}. The implementation is using the
Mondrian open source OLAP server~\cite{Mondrian} as data store.

Conceptual models are known to be very expressive and processing queries over
them is hard~\cite{CalvaneseGLLR09}. The next steps in the
materialization of the CIM Framework include the study of tractable
and yet practical cases of processing queries over our conceptual model.
A further avenue that we will pursue in the future is to relax
the assumption that the data warehouse with its schema are given and that the
only remaining task to do is to build a conceptual level over the given
data warehouse schema as well as a mapping between the conceptual level
and the warehouse schema. Figure~\ref{fig:overall_architecture2} presents
a data-centered architecture for automatic generation of the data warehouse,
assuming that only the conceptual model along with source schemas
are given and that the data warehouse with its schema (as well as
the mappings to the conceptual schema and to the sources) must be
generated and populated as automatically as possible. We intend to
expand our CIM Framework to such an architecture.

\vspace{3mm} \noindent \textbf{Acknowledgements}. We thank Daniele
Barone for his insightful comments on the manuscript. This
research was supported in part by the NSERC (Natural Sciences and
Engineering Research Council) Business Intelligence Network, IBM
Canada and SAP Business Objects division.

\bibliographystyle{alpha}
\bibliography{bibliography}

\end{document}